\documentstyle[12pt]{article}
\newcommand{\sbody}[2]{{\textstyle\frac{#1}{#2}}}
\begin{document}
\pagenumbering{arabic}
\begin{center}
\vfill
\large\bf{Aspects of the Supersymmetry Algebra}\\
\large\bf{in Four Dimensional Euclidean Space}
\end{center}
\vfill
\begin{center}
D.G.C. McKeon$^*$\\
T.N. Sherry$^{**}$\\
$^*$Department of Applied \vspace{-.5cm}Mathematics\\
University of Western \vspace{-.5cm}Ontario\\
London\vspace{-.5cm}\\
CANADA $\;\;$ N6A 5B7\\
$^{**}$Department of Mathematical \vspace{-.5cm}Physics\\
National University of \vspace{-.5cm}Ireland\\
Galway\vspace{-.5cm}\\
IRELAND
\end{center}
\vfill
email: $^*$TMLEAFS@APMATHS.UWO.CA$\;\;\;$
$^{**}$TOM.SHERRY@UCG.IE\\
Tel: (519)679-2111, ext. 8789\\
Fax: (519)661-3523
\eject

\section{Abstract}

The simplest supersymmetry (SUSY) algebra in four dimensional 
Euclidean space ($4dE$) has
been shown to closely resemble the $N = 2$ SUSY algebra in four 
dimensional Minkowski space
($4dM$). The structure of the former algebra is examined in greater 
detail in this paper. We first
present its Clifford algebra structure. This algebra shows that the
momentum Casimir invariant of physical states has an upper bound 
which is fixed by the central
charges. Secondly, we use reduction of the $N = 1$ SUSY algebra in six
dimensional Minkowski space ($6dM$) to $4dE$; this reproduces our 
SUSY algebra in $4dE$.
Moreover, this same reduction of supersymmetric Yang-Mills theory 
(SSYM) in $6dM$
reproduces Zumino's SSYM in $4dE$. We demonstrate how this 
dimensional reduction can be
used to introduce additional generators into the SUSY algebra in 
$4dE$.

\section{Introduction}

The nature of SUSY in $4dE$ is surprisingly different from that of 
SUSY in $4dM$ due to the
fact that spinors in these two spaces have distinct structures. The 
fundamental reason for this
difference is that in the decomposition of $SO(4)$ into $SU(2) \times 
SU(2)$, the generators of
the two $SU(2)$ groups are not Hermitian conjugates of each other and 
this has the consequence
that one cannot define Majorana spinors in $4dE$. A detailed analysis 
of spinors in $4dE$ and
the simplest attendant SUSY algebra is presented in [1]. There it was 
noted that this symmetry
algebra more closely resembles that of $N = 2$ SUSY in $4dM$ rather 
than $N = 1$ SUSY in
$4dM$. There are two distinct SUSY generators, their hermitian 
conjugates and two central
charges in this algebra in $4dE$; however, unlike $N = 2$ SUSY in 
$4dM$, no $SU(2)$ internal
symmetry exists between the distinct SUSY generators.

In this paper, we analyze further the algebra found in [1] in $4dE$. 
Initially, by choosing a
suitable linear combinator of SUSY generators, the Clifford algebra 
structure of this SUSY
algebra is made explicit. There is an immediate consequence of this 
algebra: the requirement that
the anticommutator of an operator and its Hermitian conjugate be 
positive definite places an
upper bound on magnitude of the eigenvalue associated with the 
Casimir $P^2$ (where $P^\mu$
is the four-momentum) that is dictated by the central charges of the 
algebra.  Furthermore, these
central charges all have to be negative definite. We note that this 
scenario is very different to
what happens in $4dM$, where the central charges can be consistently 
set to zero. In $4dE$ the
central charges must be non-zero.

The second approach to analyzing the structure of our algebra is to 
perform a dimensional
reduction of the $N = 1$ SUSY algebra from $6dM$ to $4dE$.  This is 
motivated by a similar
reduction from $6dM$ to $4dM$ done in [2,3,4]; in these references 
the $N = 1$ SSYM model
in $6dM$ is used to derive the $N = 2$ SSYM model in $4dM$. We 
actually reproduce the
$4dE$ supersymmetry algebra presented previously [1]. Surprisingly, 
by using this procedure, we
are able to extend the SUSY algebra in $4dE$ by considering 
generators initially associated with
rotation operators in $6dE$ that involve those two dimensions 
eliminated by dimensional
reduction.  We find also that by using this same reduction in 
conjunction not with the SUSY
algebra, but with the SSYM theory itself in $6dE$, the SSYM model of 
Zumino [5] in $4dE$
is automatically generated.  Furthermore, we speculate on the 
likelyhood of generating a SUSY
algebra in $4dE$ with an internal symmetry by applying dimensional 
reduction to the $N = 1$
SUSY algebra in $10dM$.

\section{The Clifford Algebra}

The simplest SUSY algebra in $4dE$ was found in [1] to be
$$\left\lbrace g, g^{c+} \right\rbrace = 0\eqno(1a)$$
$$\left\lbrace g, g^+ \right\rbrace = \gamma^\mu P^\mu + Z_+ + Z_- 
\gamma_5\eqno(1b)$$
$$\left[ P^\mu, g\right] = 0\eqno(1c)$$
$$\left[M^{\mu\nu}, g\right] = -\sbody12 \Sigma^{\mu\nu} g\eqno(1d)$$
$$\left[ M^{\mu\nu}, P^\lambda\right] = i\left(\delta^{\nu\lambda} 
P^\mu - 
\delta^{\mu\lambda} P^\nu\right)\eqno(1e)$$
$$\left[ M^{\mu\nu}, M^{\rho\sigma}\right] = i\left(\delta^{\nu\rho} 
M^{\mu\sigma} -
\delta^{\mu\rho} M^{\nu\sigma} + \delta^{\mu\sigma} M^{\nu\rho} -
\delta^{\nu\sigma}
M^{\mu\rho} \right).\eqno(1f)$$

The notation used is explained in reference [1].  We note here only 
that the charge conjugate
$g^c$ of
the Dirac spinor generator $g$ cannot be consistently set equal to 
$g$ and that $g$ cannot
decompose into a linear combination of two such (self-conjugate) 
Majorana spinors; this accounts
for the difference between (1b) and the analogous equation in $N = 2$ 
SUSY in $4dM$.

When using two dimensional notation,
$$g = \left(\begin{array}{c} Q_a \\ R^{\dot{a}}\end{array}\right)
\hspace{1in}
g^c = \left(\begin{array}{c} -{\overline{Q}}_a \\ 
{\overline{R}}^{\dot{a}}\end{array}
\right) 
\eqno(2a)$$
$$g^+ = \left({\overline{Q}}^a, -{\overline{R}}_{\dot{a}}\right) 
\qquad
g^{c+} = \left(Q^a, R_{\dot{a}}\right)\eqno(2b)$$
the algebraic expressions in (1) which involve $g$ become
$$\left\lbrace Q_a, R_{\dot{b}}\right\rbrace = 0 = 
\left\lbrace{\overline{Q}}^a,
{\overline{R}}^{\dot{b}}\right\rbrace\eqno(3a)$$
$$\left\lbrace Q_a, {\overline{R}}_{\dot{b}}\right\rbrace = i 
\sigma_{a{\dot{b}}}^\mu P^\mu
= \left\lbrace{\overline{Q}}_a, R_{\dot{b}}\right\rbrace\eqno(3b)$$
$$\left\lbrace Q_a, {\overline{Q}}_b \right\rbrace = \epsilon_{ab}
Z_{Q{\overline{Q}}}\eqno(3c)$$
$$\left\lbrace R^{\dot{a}}, {\overline{R}}^{\dot{b}} \right\rbrace =
\epsilon^{{\dot{a}}{\dot{b}}}
Z^{R{\overline{R}}}\eqno(3d)$$
where $Z_{Q{\overline{Q}}} = Z_+ + Z_-$, $Z^{R{\overline{R}}} = Z_+ - 
Z_-$.

If now we make the identifications
$$(\alpha_a, \beta_a) = (Q_a, -R^{\dot{a}})\quad (Q = 1, 2)\eqno(4a)$$
$$(\alpha_a^+, \beta_a^+) = ({\overline{Q}}^a, 
{\overline{R}}_{\dot{a}})\quad (a = 1,
2)\eqno(4b)$$
then in the frame oriented so that $P^\mu = (0, 0, 0, P)$, (3) becomes
$$\left\lbrace \alpha_a, \beta_b \right\rbrace = 0 = \left\lbrace 
\alpha_a^+, \beta_b^+
\right\rbrace\eqno(5a)$$
$$\left\lbrace \alpha_a, \beta_b^+ \right\rbrace = i \delta_{ab} P  = 
- \left\lbrace \alpha_a^+,
\beta_b\right\rbrace\eqno(5b)$$
$$\left\lbrace \alpha_a, \alpha_b^+ \right\rbrace = - \delta_{ab} 
Z_{Q{\overline{Q}}}\eqno(5c)$$
$$\left\lbrace \beta_a, \beta_b^+ \right\rbrace = - \delta_{ab} 
Z^{R{\overline{R}}}.\eqno(5d)$$
Since the left side of (5c) and (5d) are non-negative, we conclude 
that
$$Z_{Q{\overline{Q}}} \leq 0 \eqno(6a)$$
$$Z^{R{\overline{R}}} \leq 0. \eqno(6b)$$

We consider first the case where $Z_{Q{\overline{Q}}}$ and 
$Z^{R{\overline{R}}}$ are both
non-zero.  In this instance, we first make a rescaling
$$\alpha_a \longrightarrow \alpha_a\sqrt{-
Z_{Q{\overline{Q}}}}\eqno(7a)$$
$$\beta_a \longrightarrow \beta_a\sqrt{-
Z^{R{\overline{R}}}}\eqno(7b)$$
and then let
$$A_a = \frac{\alpha_a^+ - i\beta_a^+}{\sqrt{2}}\qquad A_a^+ = 
\frac{\alpha_a +
i\beta_a}{\sqrt{2}}\eqno(8a)$$
$$B_a = \frac{\beta_a + i\alpha_a}{\sqrt{2}}\qquad B_a^+ = 
\frac{\beta_a^+ -
i\alpha_a^+}{\sqrt{2}}.\eqno(8b)$$
The anticommutation relations of (5) then become
$$\left\lbrace A_a, A_b^+ \right\rbrace = \delta_{ab}(1 + \delta
I\!\!P)\eqno(9a)$$
$$\left\lbrace B_a, B_b^+ \right\rbrace = \delta_{ab}(1 - \delta 
I\!\!P)\eqno(9b)$$
(where $\delta \equiv (Z_{Q{\overline{Q}}} Z^{R{\overline{R}}})^{-
1/2}$) and all other
anticommutators involving $A_a$ and $B_a$ are zero.

For eq. (9) to be consistent, we see that we must have $(1 \pm \delta 
P) \geq 0$ so that
$$I\!\!P \leq (Z_{Q{\overline{Q}}} 
Z^{R{\overline{R}}})^{+1/2};\eqno(10)$$
this places a bound on the magnitude of the eigenvalue associated 
with the Casimir $I\!\!P^2$.

If now we were to have $Z_{Q{\overline{Q}}} = 0 = 
Z^{R{\overline{R}}}$ then it is apparent
from (5) that all states have zero norm; this case we will discard as 
uninteresting.

The last situation involves having one, but not both, of the central 
charges
$Z_{Q{\overline{Q}}}$ and $Z^{R{\overline{R}}}$  equal to zero.  
Without loss of generality,
let us consider the case $Z^{R{\overline{R}}} = 0$.  It is easily 
established that one can then
choose suitable linear combinations of $\alpha_i$, $\beta_i$, 
$\alpha_i^+$ and $\beta_i^+$ so that
$$\left\lbrace A_i, A_j^+ \right\rbrace = \delta_{ij} \left( \frac{1 
+ \sqrt{1 +
4a^2}}{2}\right)\eqno(11a)$$
$$\left\lbrace B_i, B_j^+ \right\rbrace = \delta_{ij} \left( \frac{1 -
 \sqrt{1 +
4a^2}}{2}\right)\eqno(11b)$$
(where $a \equiv I\!\!P/\sqrt{-Z_{Q{\overline{Q}}}}$) and all other 
anticommutators are zero.  For
$a^2 \neq 0$, the right side of (11b) is negative.  Since the left 
side of (11b) is positive definite,
we therefore have an inconsistency, allowing to exclude the 
possibility of having one of the
central charges equal to zero and the other non-zero.

>From the analysis of this section, we see that the SUSY algebra of 
eq. (1) is equivalent to a
Clifford algebra; the structure of this Clifford algebra imposes the 
restriction that the central
charges $Z^{R{\overline{R}}}$ and $Z_{Q{\overline{Q}}}$ be negative 
definite and that the
magnitude of the eigenvalue
of the Casimir operator $I\!\!P^2$ be always less than or equal to
$Z_{Q{\overline{Q}}}
Z^{R{\overline{R}}}$.  The saturation of this bound (so that $1 - 
\delta
I\!\!P = 0$ in (9b))
eliminate
half of the states present in the case where the bound is not 
saturated.

It is interesting to compare this situation to what occurs in $N = 2$ 
SUSY in $4dM$.  In this
latter case [6], the Clifford algebra has a structure identical to 
that of eq. (9); however the roles
of the eigenvalue of $I\!\!P^2$ and the central charges are reversed; 
 one finds that the central
charges can be zero and that there is a lower bound on the eigenvalue 
of $I\!\!P^2$ which
depends on the central charge.

We now turn to the analysis of the properties of our algebra using 
dimensional reduction of the
SUSY algebra in $6dM$ to $4dE$.

\section{Dimensional Reduction}

Most often the SUSY algebra in $4dM$ is presented using either two-
component notation or
four-component Majorana spinors; however, one can also employ four-
component Weyl spinors
to this end.  The six-dimensional analogue of this appears in [4]:
$$\left\lbrace Q, Q\right\rbrace = 0\eqno(12a)$$
$$\left\lbrace Q, {\overline{Q}}\right\rbrace = \sbody12 (1 + 
\Gamma_7)\Gamma^a
I\!\!P_a\eqno(12b)$$
$$[I\!\!M^{ab}, I\!\!P^c] = i\left(g^{bc} I\!\!P^a - g^{ac} 
I\!\!P^b\right)\eqno(12c)$$
$$[I\!\!M^{ab}, I\!\!M^{cd}] = i\left(g^{bc} I\!\!M^{ad} - g^{ac} 
I\!\!M^{bd} + g^{ad}
I\!\!M^{bc} - g^{bd} I\!\!M^{ac} \right)\eqno(12d)$$
$$[I\!\!M^{ab}, Q] = -\sbody12 \Sigma^{ab} Q.\eqno(12e)$$
Here, we have taken $Q$ to be an eight component Dirac spinor with 
$\overline{Q} = Q^+
\Gamma^0$.

The Dirac matrices are taken to be
$$\Gamma^0 = \left( \begin{array}{cc} 0 & -i \\ i & 0 \end{array} 
\right)\;\;\;\;
\Gamma^i = \left( \begin{array}{cc} 0 & i\gamma^i \\ i\gamma^i & 0 
\end{array}
\right) \;\;\; (i = 1, 2, 3)\nonumber$$
$$\Gamma^5 = \left( \begin{array}{cc} 0 & i\gamma_5 \\ i\gamma_5 & 0
\end{array} \right)\;\;\;\;
\Gamma^6 = \left( \begin{array}{cc} 0 & i\gamma^0 \\ i\gamma^0 & 0 
\end{array}
\right) \eqno(13)$$
$$\Sigma^{ab} = \frac{i}{2} [\Gamma^a, \Gamma^b]\nonumber$$
$$\Gamma_7 = \Gamma^0 \Gamma^1 \Gamma^2 \Gamma^3 \Gamma^5 \Gamma^6 = 
\left(
\begin{array}{cc} -1 & 0 \\ 0 & 1\end{array}\right)\nonumber$$

where $\left\lbrace\gamma^\mu, \gamma^\nu \right\rbrace = 
2\delta^{\mu\nu}$
$(\mu,\nu = 1 \ldots 4)$ and $\gamma_5^+
= \gamma_5$.

      In order to reduce the algebra of (12) to that of (1), we make 
the identification
$$Q = \left( \begin{array}{c} 0 \\ g \end{array} \right)\eqno(14)$$
where $g$ is a four component spinor, and define,
$$I\!\!P_0 = Z_+ \eqno(15a)$$
$$I\!\!P_5 = Z_- \eqno(15b)$$
$$(I\!\!P_1, I\!\!P_2, I\!\!P_3, I\!\!P_6) = P^\mu\eqno(15c)$$
$$M^{ij} = - I\!\!M^{ij}\;\; (i,j = 1, 2, 3)\eqno(15d)$$
$$M^{i4} = -I\!\!M^{i6}.\eqno(15e)$$

It is easily verified that with these identifications, the algebra of 
(12) reduces to that of (1).

The role of the rotation operators in (12) merits consideration.  We 
can define
$$J^\mu = I\!\!M^{\mu 0}\eqno(16a)$$
$$K^\mu = I\!\!M^{\mu 5}\eqno(16a)$$
$$L = I\!\!M^{05}\eqno(16c)$$
and obtain from (12) the non-zero commutators (which are consistent 
with those of (1), in the
sense that all Jacobi identities are satisfied):
$$[L,g] = \frac{i}{2} \gamma_5 g\eqno(17a)$$
$$[L, Z_{\pm} ] = iZ_{\mp}.\eqno(17b)$$
$$[J^\mu, g] = -\frac{i}{2} \gamma^\mu g\eqno(17c)$$
$$[K^\mu, g] = \frac{i}{2} \gamma^\mu \gamma_5 g\eqno(17d)$$
$$[J^\mu, P^\nu] = -i \delta^{\mu\nu} Z_+\eqno(17e)$$
$$[K^\mu, P^\nu] = i \delta^{\mu\nu} Z_-\eqno(17f)$$
$$[J^\mu, Z_+] = -i P^\mu \eqno(17g)$$
$$[K^\mu, Z_-] = -i P^\mu\eqno(17h)$$
$$[J^\mu, L] = iK^\mu\eqno(17i)$$
$$[K^\mu, L] = iJ^\mu.\eqno(17j)$$
$$[J^\mu, K^\nu] = i \delta^{\mu\nu} L\eqno(17k)$$
$$[J^\mu, J^\nu] = i M^{\mu\nu} = - [K^\mu, K^\nu]\eqno(17l)$$

Together, (1) and (17) constitute an algebra which is an extension of 
the algebra presented in [1]
by itself.  It is evident that the $N = 2$ SUSY algebra in $4dM$ can 
be extended in a similar
fashion by dimensionally reducing the $6dM$ algebra of (12), as 
outlined in ref. [4].

It is also possible to perform a dimensional reduction of the $N = 1$ 
SUSY algebra in $10dM$
to $4dM$ to obtain the $N = 4$ SUSY algebra in $4dM$.  It is likely 
that an extended SUSY
algebra in $4dE$ can be generated by a similar dimensional reduction.

Dimensional reduction was primarily used in ref. [2-3] to generate a 
SSYM model with extended
SUSY in $4dM$.  It is interesting to note at this point that the SUSY 
model of Zumino [5] can
similarly be generated.  One starts with the action in $6dM$
$$I = \int d^6x \left[ -\sbody14 F_{ab} F^{ab} + \sbody12 
{\overline{\lambda}} \Gamma^a
{\stackrel{\leftrightarrow}{\nabla}}_a \lambda \right]\eqno(18)$$
with
$$\lambda = \sbody12 \left(1 - \Gamma_7 \right)\lambda\nonumber$$
$$\nabla_a = \partial_a \lambda + i [A_a, \lambda].\nonumber$$
To dimensionally reduce this action to $4dE$, one employs the 
representation of the
$\Gamma^a$
given in (13), sets $\partial_0 = \partial_5 = 0$ and makes the 
identifications
$$\lambda = \left( \begin{array}{c} \psi \\ 0 \end{array} 
\right)\eqno(19)$$
$${\overline{\lambda}} = (0, -i\psi^+)\nonumber$$
$$A_0 = A\nonumber$$
$$A_5 = B\nonumber$$
in order to produce the Zumino model.  (One could presumably deduce a 
SSYM with extended
supersymmetry in $4dE$ by a dimensional reduction of the $10dM$ 
version of the model of eq.
(18).)  We should keep in mind that the momentum of ``physical'' 
states in the Zumino model
are bounded by the inequality of (10).
\section{Discussion}

In this paper we have considered two aspects of the SUSY algebra in 
$4dE$ introduced in [1]. 
First of all, the algebra has been written as a Clifford algebra.  
This has demonstrated that the
algebra is closely linked to that of $N = 2$ SUSY algebra in $4dM$, 
the principle difference
being the inversion of the roles of momentum and central change so 
that the eigenvalue of the
Casimir operator $P^2$ faces an upper bound determined by the central 
charges.  Secondly, we
have generated both the SUSY algebra in $4dE$ and the Zumino model of 
ref. [5] by
dimensional reduction from $6dM$.  This procedure has shown how 
additional
Bosonic operators
$J^\mu$, $K^\mu$ and $L$ can be introduced, as in (17), thereby 
extending the algebra of (1). 
This is unexpected in view of the theorems in ref. [7].  We also note 
that dimensional reduction
was also used in ref. [8] to study spinors in $4dE$.

\section{Acknowledgement}
 
The authors would like to thank each other's home institutions for 
hospitality while this work
was being done.  Financial support was provided by NSERC and Forbairt.

\end{document}